\documentclass[twocolumn,superscriptaddress,amsmath,amssymb,aps,prl,floatfix]{revtex4-2}
\usepackage[usenames,dvipsnames]{color}
\usepackage{xcolor}
\usepackage{multibib} 

\usepackage{comment}
\usepackage{cancel}
\usepackage{ulem}
\usepackage{float}
\usepackage{physics}
\usepackage{graphicx}
\usepackage{dcolumn}
\usepackage{multirow}
\usepackage{bm}
\usepackage{gensymb}
\usepackage{tensor}
\usepackage[percent]{overpic}
\usepackage{tikz}
\usepackage{pgfplots}
\pgfplotsset{compat=1.15}
\usepackage{mathrsfs}
\usepackage{multibib}

\usepackage[breaklinks=true,colorlinks,citecolor=blue,linkcolor=blue,urlcolor=blue]{hyperref}

\footnotetext{These authors contributed equally to this work.}
\footnotetext{francesco.cioni@sns.it}
\begin{document}
\title{Persistent currents, whirlpools, and local Chern markers in twisted TMD  Chern insulators}

\author{Francesco Cioni$^\star$${}^\ddag$}
\affiliation{NEST, Scuola Normale Superiore, I-56126 Pisa,~Italy}
\author{Lorenzo Cavicchi$^\star$}
\affiliation{Scuola Normale Superiore, Piazza dei Cavalieri 7, I-56126 Pisa,~Italy}
\author{Nazzareno Africani}
\affiliation{International School for Advanced Studies (SISSA), Via Bonomea 265, Trieste 34136, Italy}
\author{Giacomo Mazza}
\affiliation{Dipartimento di Fisica dell'Universit\`a di Pisa, Largo Bruno Pontecorvo 3, I-56127 Pisa,~Italy}
\author{Fabio Taddei}
\affiliation{Istituto Nanoscienze – CNR, NEST-SNS, Piazza San Silvestro 12, I-56127 Pisa,~Italy}
\author{Amir Yacoby}
\affiliation{Department of Physics, Harvard University, Cambridge, Massachusetts 02138, USA}
\author{Marco Polini}
\affiliation{Dipartimento di Fisica dell'Universit\`a di Pisa, Largo Bruno Pontecorvo 3, I-56127 Pisa,~Italy}
\begin{abstract}
Recent materials advances have made it possible to fabricate twisted transition metal dichalcogenide homobilayers. These systems have been shown to host integer and fractional Chern insulating states. Because of spontaneous time reversal symmetry breaking, their ground state harbors intriguing spin-polarized currents with whirlpools on the moir\'e length scale that can be measured by scanning probe methods. We first provide a quantitative analysis of these persistent currents and then show that the maximum of the amplitude of the current density in the bulk of the sample is an accurate tracker of topological order. We conclude by calculating how the quantization of the Hall conductance is affected by finite-size effects. 
\end{abstract}
\maketitle

{\color{blue} \it Introduction.}---Topological phases in crystalline materials are characterized by invariants, such as the Chern number ${\cal C}$, defined as global quantities integrated over the entire Brillouin zone~\cite{Thouless_PRL_1982, Haldane_PRL_1988, vanderbilt, bernevig}. The Chern number, in particular, directly determines the existence of topologically protected boundary states, present within a bulk energy gap, through the principle of the so-called ``bulk–boundary'' correspondence~\cite{Halperin_1982, Hatsugai_PRL_1993, Kane_2005}. Such boundary modes, which are a genuine signature of the underlying topology, are protected against elastic perturbations that do not close the bulk energy gap~\cite{Xiao_RMP_2010, Nagaosa_RMP_2010} (although they are fragile against inelastic scattering---see e.g. Ref.~\cite{novelli_prl_2019} and references therein to earlier work). 
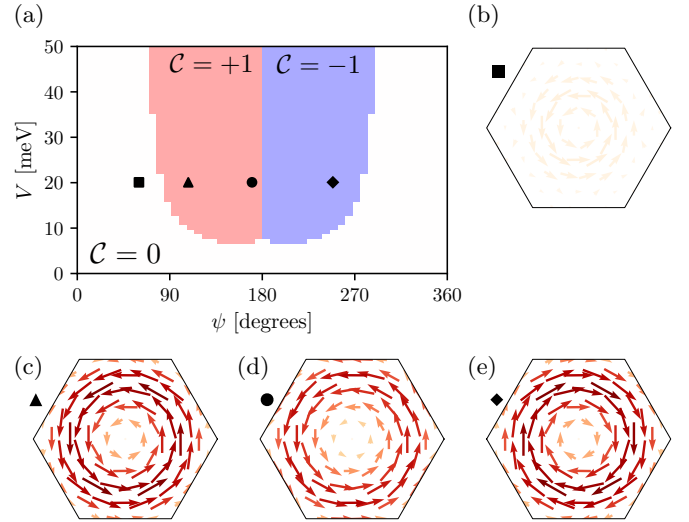
\begin{figure}
    \centering
    \begin{overpic}
        [width = 0.5\textwidth]{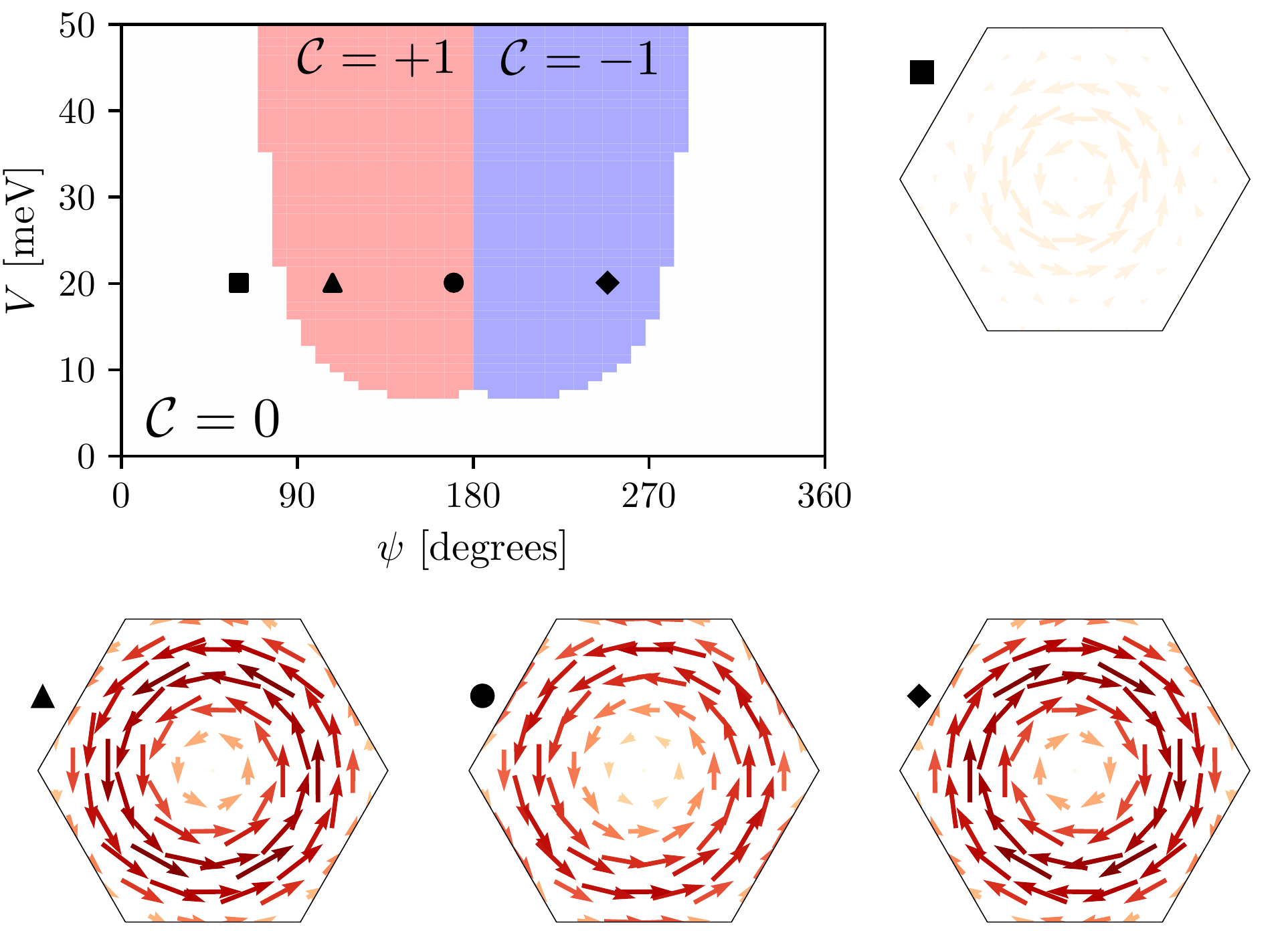}
        \put(0,74){(a)}
        \put(67,74){(b)}
        \put(0,22){(c)}
        \put(33,22){(d)}
        \put(67,22){(e)}
    \end{overpic}
    \caption{Topological phase diagram of the skyrmion Chern-band continuum model~\cite{Wu2019,morales-duran_PRL_2024}---panel (a)---as a function of the two  parameters $V$ and $\psi$ in Eq.~\eqref{delta_ell}. White regions indicate topologically trivial phases while colored regions denote phases with ${\cal C}\neq 0$. The triangle corresponds to twisted MoTe$_2$.  We also show the current whirlpools circulating inside the moir\'e unit cells for four different choices of parameters. In the trivial phase---square, panel (b)---the ground-state current has a very small amplitude, while it becomes large as soon as one enters into the topological phase, panels (c) to (e). Notice the chirality reversal in passing from positive (black triangle) to negative values of ${\cal C}$ (rhomboid).}
    \label{fig:prima_figura}
\end{figure}

More recently, Resta and co-workers introduced the concept of ``local'' topological markers~\cite{bianco_PRB_2011, Bianco_PRL_2013, Marrazzo_PRB_2017, Marrazzo_PRL_2019}. Key advantage of these {\it real-space} markers is that they are able to identify the topological and geometrical nature of a system even in those cases in which a Brillouin zone cannot be defined, as in the case of disordered materials~\cite{Bau_PRB_2024, Favata_PRL_2025}. Moreover, it has been shown that local Chern markers can be used to extract the critical behavior of topological phase transitions~\cite{caio_NATPHYS_2019}. Contrary to the Chern number (which is directly related to the Hall conductivity~\cite{Thouless_PRL_1982} and therefore readily measurable), however, the real-space topological markers that have been proposed so far~\cite{bianco_PRB_2011, Bianco_PRL_2013, Marrazzo_PRB_2017, Marrazzo_PRL_2019, caio_NATPHYS_2019} cannot be measured experimentally.

This naturally leads to the following key question: Is there a measurable local quantity acting as a real-space marker for topological phases with non-zero Chern number? Concentrating on the case of two-dimensional (2D) systems, in this work we show that the answer to this question is affirmative and deceptively simple. Systems with a non-zero Chern number break time-reversal symmetry and are therefore expected, on general grounds, to display an interesting spatial distribution ${\bm J}({\bm r})$ of {\it ground-state} currents. However, as stated above, conventional wisdom is centered on current flow near the sample edges. Here, we look more generally at the full real-space distribution of ground-state currents, both in the bulk and near the edges of finite-size samples. We find that the maximum of the amplitude of the ground-state current density in bulk of the sample, i.e.~$\max_{{\bm r} \in {\rm bulk}}|{\bm J}({\bm r})|$, is a rather accurate local marker of Chern insulating states. We carry out detailed numerical calculations for a moir\'{e} 2D quantum material, where integer and fractional Chern insulating states have been experimentally discovered. These are twisted transition‑metal dichalcogenide (TMD) homobilayers, which are well described by skyrmion Chern-band continuum models~\cite{Wu2019,morales-duran_PRL_2024,pan_2020,reddy_prl_2024,shi_PRB_2024,cavicchi_npjQuantum_2025}. 

Twisted TMD homobilayers host narrow, topological moir\'e bands in which the interplay between strong electronic correlations and quantum geometry produces incredibly rich phase diagrams, ranging from ferromagnetism~\cite{Devakul2021} to fractional Chern insulators~\cite{reddy2023, Duran2024, wang2024} and superconductivity~\cite{guerci2025, Qin2025}.
In particular, twisted TMD homobilayers host time-reversed topological Chern bands at the $K$ and $K'$ valleys, ``locked'' to the spin degree of freedom by the strong spin–orbit coupling inherited from TMD monolayers~\cite{Xiao2012}. Experiments in twisted bilayer MoTe$_2$ have provided robust evidence for the existence of integer and fractional Chern insulating states at zero magnetic field~\cite{Cai2023,Park2023,Xu2023,Zeng2023}, with spontaneous breaking of time-reversal symmetry (TRS) at integer filling $\nu=-1$ and at several Jain-sequence fractional fillings, most prominently $\nu=-2/3$ and $\nu=-3/5$~\cite{Cai2023,Park2023,Zeng2023}, with subsequent local magnetic imaging reporting signatures also at $\nu=-4/7$ and $\nu=-5/9$~\cite{redekop_2024,wang_2025}. Very recently, superconductivity has been observed in twisted bilayer ${\rm WSe}_2$ close to $\nu=-1$~\cite{Xia2025, Guo2025}. 

Because these correlated Chern phases spontaneously break TRS, their ground states carry orbital magnetization and chiral currents, even in the absence of an applied magnetic field. Imaging via scanning SQUID magnetometry~\cite{Tschirhart2021Science, Zhou2023} of moir\'e Chern insulators in related graphene-based platforms has already revealed micron-scale ferromagnetic domains, pseudo-magnetic fields, and chiral edge currents. In moir\'e materials, scanning NV and other local probes have directly visualized spatially modulated (moir\'e-scale) magnetism~\cite{casola_2018_main, Song2021Science}. Hydrodynamic flow~\cite{viscous_flow_review, ku_2020, Vool_2021} and superconducting currents~\cite{Chen_2024} have also been observed using these techniques.

Motivated by these technical developments in scanning probe methods, we analyze the spatial dependence of the ground-state current density ${\bm J}({\bm r})$ in twisted TMD homobilayers ribbons, with particular attention to the patterns occurring on the moir\'e length scale~\cite{foot_magnetization}.
We demonstrate that such local currents, measured in the bulk of the ribbon, provide an effective probe of the topological phase. Indeed, pronounced current whirlpools emerge within bulk unit cells (BUCs) only when the system is in the topological regime.
This is summarized in Fig.~\ref{fig:prima_figura}, which presents the topological phase diagram (TPD) of the skyrmion Chern-band continuum model~\cite{Wu2019,morales-duran_PRL_2024}---panel (a)---together with the calculated intra-cell current density patterns corresponding to four points in the TPD. The current density is finite in the three topological cases with a non-vanishing Chern number ${\cal C}\neq 0$: in the region with ${\cal C}=-1$ (filled rhomboid), in the region with ${\cal C}=+1$ (filled triangle), and also in close proximity to the topological phase boundary between ${\cal C}=+1$ and ${\cal C}=-1$ (filled circle). On the contrary, the current density is vanishingly small (i.e.~up to 20 times smaller than in panels (c) to (e)) in the trivial phase (black square)---see panel (b). Remarkably, we find that the maximum amplitude of the current density, evaluated by scanning ${\bm r}$ over a BUC located at the center of the sample,
\begin{equation}\label{eq:novel_quantifier}
{\cal J}\equiv \max_{{\bm r} \in {\rm BUC}}|{\bm J}({\bm r})|~,
\end{equation}
is a reliable local proxy of topological phases with ${\cal C}\neq 0$. By using the Biot-Savart law, we also estimate the magnetic fields generated by such ground-state currents, discussing the feasibility of measuring them via state-of-the-art scanning probe methods. Finally, we investigate the effects of edge hybridization in finite-size ribbons (see End Matter) by computing how the transverse Hall conductivity $\sigma_{xy}$ approaches its bulk value ${\cal C} e^2/h$, extracting two ``critical'' exponents from a finite-size scaling analysis.

{\color{blue} \it Model Hamiltonian.}---We examine a generic twisted TMD homobilayer formed by stacking two TMD monolayers, separated by a distance $d$, and rotating one layer relative to the other by a twist angle $\theta$. Low-energy valence-band electrons are described by the matrix Hamiltonian
\begin{equation}\label{eqn:H_tot}
    \mathcal{H} = \begin{pmatrix} \mathcal{H}_{\uparrow} & 0 \\
    0 & \mathcal{H}_{\downarrow}\end{pmatrix}~,
\end{equation}
where $\mathcal{H}_{\uparrow}$ and $\mathcal{H}_{\downarrow}$ are continuum-model Hamiltonians relative to spin-up and spin-down electrons, respectively (or, equivalently, to the $+K$ and $-K$ valleys, due to spin-valley locking~\cite{Xiao2012}). The spin-up/$K$-valley Hamiltonian 
can be written in the form~\cite{Wu2019}
\begin{equation}
\label{H_up}
    \mathcal{H}_\uparrow = \begin{pmatrix} -\frac{\hbar^2(\bm{k} - \bm{k}_{\rm b})^2}{2m^*} + \Delta_{\rm b}(\bm{r}) & \Delta_{\rm T}(\bm{r}) \\ \Delta_{\rm T}^*(\bm{r}) & -\frac{\hbar^2(\bm{k} - \bm{k}_{\rm t})^2}{2m^*} + \Delta_{\rm t}(\bm{r}) \end{pmatrix}~,
\end{equation}
and $\mathcal{H}_\downarrow$ is obtained by applying the time-reversal operation to $\mathcal{H}_\uparrow$.
The Hamiltonian in Eq.~\eqref{H_up} is represented in the layer-pseudospin space, where the label ${\rm b}$ (${\rm t}$) refers to the bottom (top) layer.
The momentum shifts $\bm{k}_{\rm b/t}$ correspond to the positions of the $+K$ valleys of the top and bottom layers, respectively, after rotation, namely
\begin{equation}\label{eqn:K_ell}
    \bm{k}_{\rm b/t} = \frac{4\pi}{3a_{\rm M}}\left(\pm \frac{1}{2},  \frac{\sqrt{3}}{2}\right)~,
\end{equation}
where the plus (minus) sign refers to the top (bottom) layer.
Here, $a_{\rm M} = a/[2\sin(\theta/2)]$
is the moir\'e lattice constant, with $a$ as the lattice parameter of the underlying monolayer TMD. The inter-layer potential $\Delta_{\rm T}$ is defined as
\begin{equation}\label{delta_T}
    \Delta_{\rm T}({\bm{r}}) \equiv w \left[1 + e^{i\bm{G}_2(\theta)\cdot\bm{r}} + e^{i\bm{G}_3(\theta)\cdot\bm{r}}\right] ,
\end{equation}
while the intra-layer potentials are
\begin{equation}\label{delta_ell}
    \Delta_{\rm b/t}(\bm{r}) \equiv 2V\sum_{j=1,3,5}\cos\left[\bm{G}_j(\theta)\cdot\bm{r}\pm\psi\right]~,
\end{equation}
where the plus (minus) sign refers to the top (bottom) layer.
The intra-layer potentials, exhibiting periodicity over the moir\'e unit cell, originate from the differing alignments between the metal and chalcogen atoms in the top and bottom TMD monolayers~\cite{Wu2019}. The parameters $V$, $\psi$, and $w$ specify the TMD monolayer material. For example, for $\mathrm{MoTe}_2$, we have~\cite{wang2024} $V=20.8$ meV, $\psi=107.7\degree$, and $w=-23.8$ meV. In Eqs.~(\ref{delta_T}) and~(\ref{delta_ell}), $\bm{G}_j(\theta)$ are the reciprocal lattice vectors belonging to the first shell of the moir\'e reciprocal lattice and are defined by 
\begin{equation}
\bm{G}_j(\theta) = \frac{4\pi}{\sqrt{3}a_{\rm M}}\left(-\sin\left[(j-1)\frac{\pi}{3}\right], \cos\left[(j-1)\frac{\pi}{3}\right]\right)~.
\end{equation}

As shown in Fig.~\ref{fig:prima_figura}, the Hamiltonian in Eq.~\eqref{H_up} possesses two different topological phases, in addition to a trivial phase, depending on the values of its microscopic parameters $V$, $\psi$, and $w$. For example, for $\theta = 3.89\degree$, a calculation of the TPD shows that by fixing $w$ and $V$ such that $V/w\gtrsim 1/4$, and increasing $\psi$ from zero to $360\degree$, the Chern number ${\cal C}$ of the topmost valence band goes from 0 to 1 at $\psi \approx 85\degree$,
while it jumps from 1 to $-1$ at $\psi = 180\degree$. It finally returns to 0 for $\psi\gtrsim 275\degree$.
From a linear-response perspective, a non-zero Chern number results in finite Hall conductivity, which is due to the presence of dispersive states localized on the edges of the ribbon~\cite{vanderbilt}.

{\color{blue} \it Ribbon geometry.}--- In what follows, we study a ribbon of a twisted TMD homobilayer, which is infinite in the $x$-direction and finite in the $y$-direction, with a width $W$.
Since the potentials $\Delta_{\rm T}({\bm{r}})$ and $\Delta_{\rm b,t}({\bm{r}})$, defined in Eqs.~\eqref{delta_T} and~\eqref{delta_ell}, respectively, have the periodicity of the moir\'e lattice, we can apply Bloch's theorem along the $x$-direction. Strong electron-electron interactions in these materials lift valley and spin degeneracies, favoring a flavor-polarized ground state that breaks TRS~\cite{wang2024,reddy2023,yu_PRB_2024}. Accordingly, in what follows, we focus on the single-spin, single-valley Hamiltonian in Eq.~\eqref{H_up}.  The eigenstates of this Hamiltonian are thus labeled by two quantum numbers: a band index $n$ and a wave vector $k_x$, which takes values in the one-dimensional Brillouin zone $[-\pi/a_{\rm M}, \pi/a_{\rm M}]$. The band index $n$ includes the transverse modes due to the finite-width $W$ of the ribbon along the $y$-direction. The eigenfunctions can be written in the form
\begin{equation}\label{eqn:bloch:eigfun}
    \langle \bm{r} \vert k_x, n \rangle \equiv  \langle \bm{r} \vert\begin{pmatrix}
        \vert k_x, n, {\rm b} \rangle \\ \vert k_x, n, {\rm t} \rangle
    \end{pmatrix} = e^{ik_x x} \begin{pmatrix}
        u^{({\rm b})}_{k_x,n}(x,y) \\ u^{({\rm t})}_{k_x,n}(x,y)
    \end{pmatrix}~, 
\end{equation}
where the functions $u^{({\rm b})}_{k_x,n}(x,y)$ and $u^{({\rm t})}_{k_x,n}(x,y)$ are periodic in the $x$-direction and vanish at $y = 0,\,W$. In order to find the eigenfunctions~\eqref{eqn:bloch:eigfun} and the associated eigenenergies $\varepsilon_{k_x,n}$ of the system, we solved the quasi-1D Schroedinger equation $\langle \bm{r}\vert \mathcal{H}_{\uparrow} \vert k_x, n \rangle = \varepsilon_{k_x,n} \langle \bm{r} \vert k_x, n \rangle$ as detailed in Sect.~\ref{eq:Schroed_ribbon_geometry} of the Supplemental Material (SM)~\cite{SM}. 
The Hamiltonian ${\cal H}_{\uparrow}$ in a ribbon geometry has been recently studied in Ref.~\cite{Saleem_2026}, where the authors calculated the electron {\it density} of the edge modes. 

\begin{figure*}
    \centering
    \begin{overpic}
        [width = \textwidth]{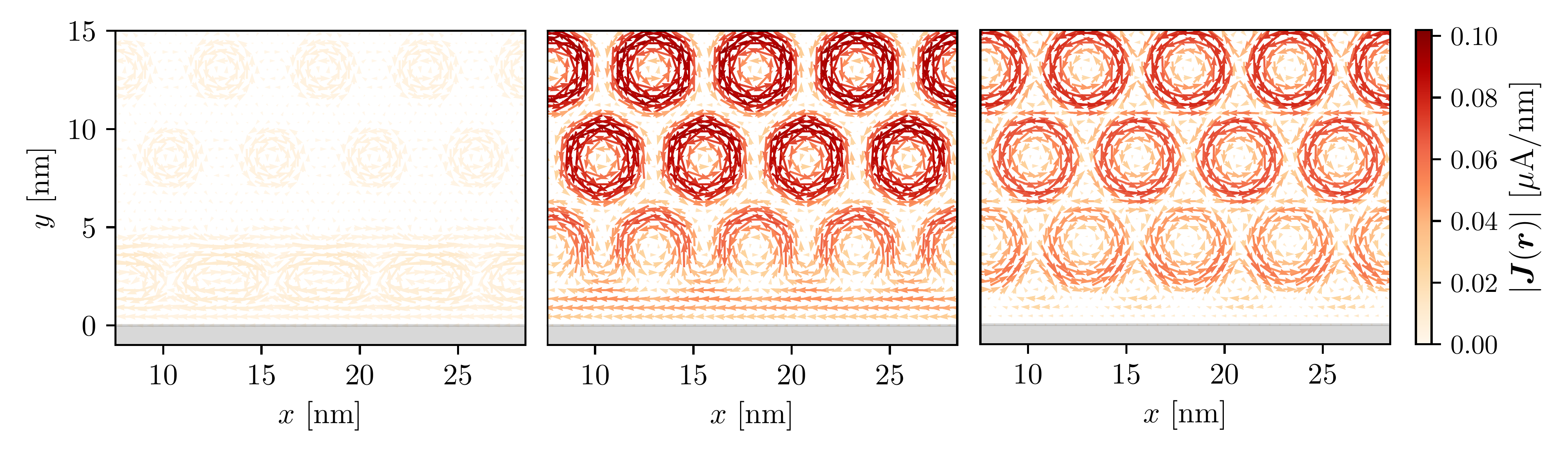}
        \put(5,29){(a)}
        \put(34,29){(b)}
        \put(62,29){(c)}
    \end{overpic}
    \caption{Ground-state currents near the bottom edge of a ribbon of width $W \simeq 72~{\rm nm}$ of a twisted TMD homobilayer with a twist angle $\theta = 3.89 \degree$. The area shaded in gray represents the edge of the ribbon. The figures depict the local current density $\bm{J}(\bm{r})$ as defined in Eq.~\eqref{eq:total_j} for three values of the microscopic parameter $\psi$ corresponding to the upper band being in the: (a) trivial ($\psi = 60\degree$) phase, (b) {\it near-critical} ($\psi = 107.7\degree$, representing twisted MoTe$_2$) phase, and (c) in the topological phase ($\psi = 170\degree$).  The direction of the arrows corresponds to the direction of the vector field $\bm{J}(\bm{r})$ while their color indicates the intensity. The results in this figure have been obtained by setting $w=-23.8~{\rm meV}$ and $V=20.8~{\rm meV}$. The value of $W$ has been set at $W=8\sqrt{3}a_{\rm M}$, where the moir\'{e} length for $\theta = 3.89 \degree$ is $a_{\rm M} \simeq 5.2\ {\rm nm}$.
    \label{fig:gs_currents}}
\end{figure*}

Here, we emphasize that, since TRS is spontaneously broken, the ground-state of ${\cal H}_{\uparrow}$ is expected to host equilibrium, ground-state {\it currents}. We now set out to study the properties of such currents.

{\color{blue} \it Real-space current density pattern.}---The single-particle local current density operator is defined as $\hat{\bm{\jmath}}(\bm{r}) = -c\delta \hat{\mathcal{H}}_\uparrow(\bm{A}(\hat{\bm{r}}))/\delta \bm{A}(\hat{\bm{r}})$, where $\hat{\mathcal{H}}_\uparrow(\bm{A}(\hat{\bm{r}}))$ is the Hamiltonian in Eq.~\eqref{H_up}, minimally coupled to a classical vector potential $\bm{A}(\hat{\bm{r}})$. By computing the functional derivative (see Sect.~\ref{sect:currop_SM} of the SM~\cite{SM}) we find that the operator $\hat{\bm{\jmath}}(\bm{r})$ is diagonal in the layer degree of freedom:
\begin{equation}
\label{jj}
    \hat{\bm{\jmath}}(\bm{r})  = \begin{pmatrix} \hat{\bm{\jmath}}^{\rm (b)}(\bm{r}) & 0 \\ 0 & \hat{\bm{\jmath}}^{\rm (t)}(\bm{r}) \end{pmatrix}~.
\end{equation}
Therefore, the total ground-state current density is obtained by summing over the occupied states:
\begin{equation}\label{eq:total_j}
    \bm{J}(\bm{r}) = \sum_{(k_x,n) \in \rm occ} \sum_{\gamma = \rm b,t} \langle k_x, n, \gamma \vert \ \hat{\bm{\jmath}}^{(\gamma)}(\bm{r}) \ \vert k_x,n, \gamma \rangle~,
\end{equation}
where the quantity $\langle k_x, n, \gamma \vert \ \hat{\bm{\jmath}}^{(\gamma)}(\bm{r}) \ \vert k_x,n, \gamma \rangle$ (for $\gamma ={\rm b}, {\rm t}$) is the expectation value of the one-body operator $\hat{\bm{\jmath}}^{(\gamma)}(\bm{r})$: 
\begin{widetext}
\begin{eqnarray}\label{current_layer}
    \langle k_x, n, \gamma \vert \ \hat{\bm{\jmath}}^{(\gamma)}(\bm{r}) \ \vert k_x,n, \gamma \rangle =  \frac{e\hbar}{2m^*} \Big [i(\nabla u^{(\gamma)}_{k_x,n}(\bm{r}))^* u^{(\gamma)}_{k_x,n}(\bm{r}) - i(u^{(\gamma)}_{k_x,n}(\bm{r}))^* \nabla u^{(\gamma)}_{k_x,n}(\bm{r}) \Big] + \frac{e\hbar}{m^*} \Big[ k_x \hat{\bm{x}} - \bm{k}_\gamma\Big ] |u^{(\gamma)}_{k_x,n}(\bm{r})|^2 ,
\end{eqnarray}
\end{widetext}
which is periodic along the $x$ direction due to the periodicity of the functions $u^{(\gamma)}_{k_x,n}(\bm{r})$. Note that, since twisted TMDs inherit spin-valley locking from the constituent monolayers~\cite{Xiao2012}, the ground-state current density ${\bm J}({\bm r})$ is spin polarized.

Our aim is now to assess whether the local, ground-state current density ${\bm J}({\bm r})$ is sensitive to the topological properties of the bands. For the sake of concreteness, we focus on a twist angle $\theta = 3.89\degree$. Figure~\ref{fig:gs_currents} shows $\bm{J}(\bm{r})$---represented by the small arrows---and its modulus---represented by the colorbar---in a portion of the ribbon close to the bottom edge, for three values of the parameter $\psi$, namely $60\degree$ [panel (a)], $107.7\degree$ [panel (b)], and $170\degree$ [panel (c)], assuming fixed $w=-23.8~{\rm meV}$ and $V=20.8~{\rm meV}$. These three values of $\psi$ correspond to the upper band being, respectively, in the trivial phase (${\cal C}=0$), in the {\it near-critical} phase (with ${\cal C} =1$, but near the topological phase boundary), and deep in the topological phase (${\cal C}=1$). These three configurations correspond to the black square, black triangle, and black circle depicted in Fig.~\ref{fig:prima_figura}, respectively.

Fig.~\ref{fig:gs_currents} reveals a rich structure of local current patterns emerging in the ground state of the system, from which we identify three main features.
First, regions in which the current exhibits a circular-like rotation about a central point are denoted as {\it whirlpools}.  Second, near the system edge [see in particular Fig.~\ref{fig:gs_currents}(b)], we observe truncated circular trajectories that morph into a current with a preferred direction of motion; such patterns are referred to as {\it skipping orbits}. Finally, whenever the local current shows, either locally or on average, a net left- or right-moving flow along the edge, we refer to it as an {\it edge current}.
In the following, we analyze whether these features are related to the  topological properties of the system.

In the trivial phase, Fig.~\ref{fig:gs_currents}(a), the current density remains very small throughout the sample, exhibiting in the bulk only faint signs of whirlpool-like features, which resemble cyclotron orbits.
A small, locally-finite edge current can be recognized, but a closer look at the pattern reveals that the average current at the edge is approximately zero due to the coexistence of both left- and right-moving flows (see also the discussion below). The weakness of bulk currents in the ${\cal C}=0$ phase is somewhat unexpected, given that TRS is broken.

On the other hand, pronounced current whirlpools are evident in both Figs. \ref{fig:gs_currents}(b) and (c), i.e., when the system is in the topological phase. 
Although the bulk current density behaves similarly in these two cases, their behavior at the ribbon’s edge is remarkably different.
Indeed, we find that strong edge currents reminiscent of skipping orbits are present only in the near-critical phase [panel (b)] but are absent deep in the topological phase [panel (c)]. Moreover, in the latter case, a strong edge current with the opposite chirality compared to the
case in panel (b)
is observed~\cite{footnote_2}.

\begin{figure}
    \centering
    \begin{overpic}[width=0.9\linewidth]{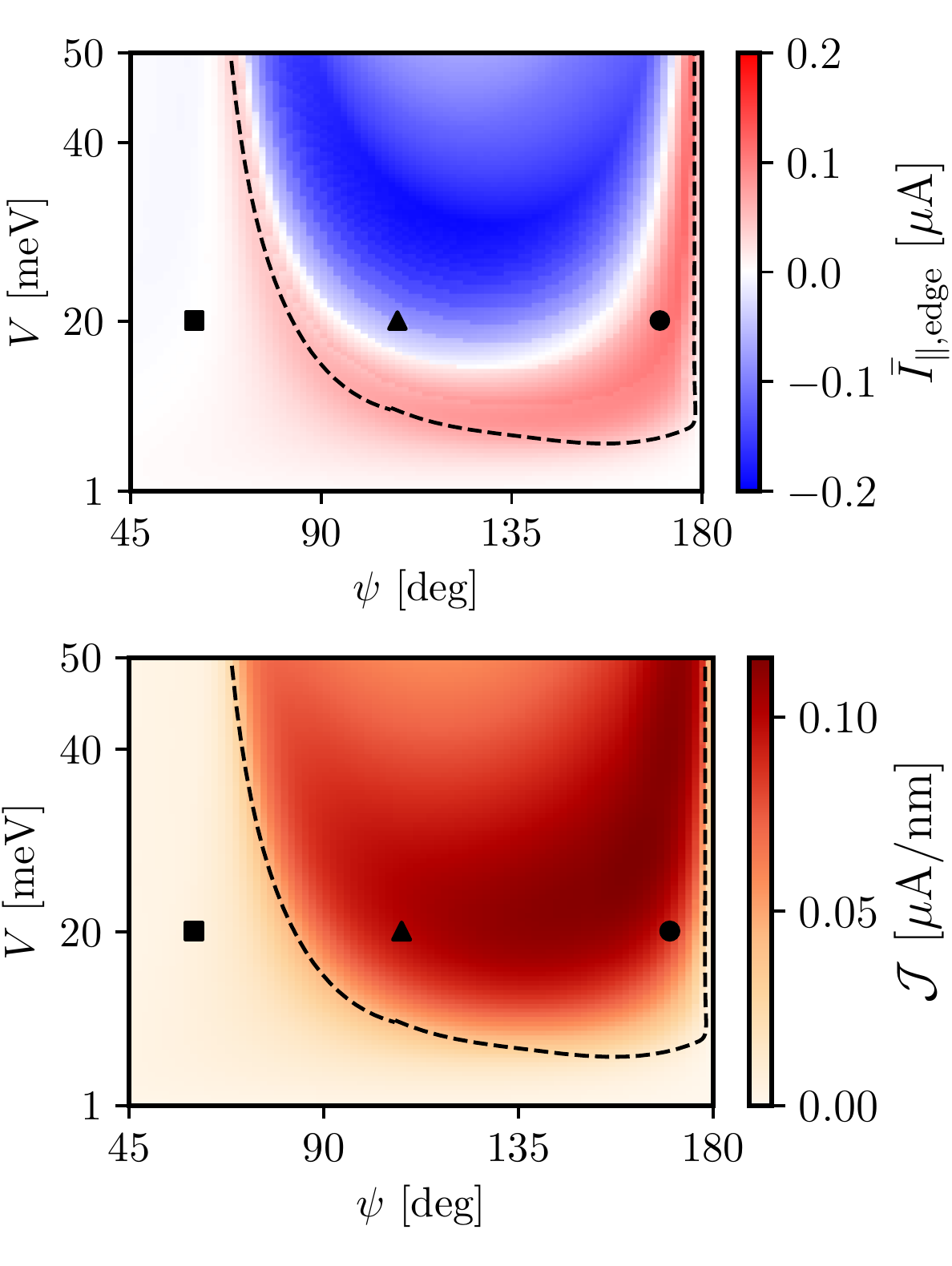}
        \put(-5,96){(a)}
        \put(-5,46){(b)}
    \end{overpic}
    
    \caption{Comparison between edge and bulk ground-state currents in a twisted TMD ribbon. In both panels, the black dashed line identifies the topological phase transition. The square, triangle and circle correspond, respectively, to the three configurations in panel (a), (b), and (c) of Fig.~\ref{fig:gs_currents}, respectively. Panel (a) shows the average current (\ref{eq:average_edge_current}) (in $\mu{\rm A}$) flowing in the $x$-direction near the bottom edge of the ribbon, i.e.~for $y \in [0, \sqrt{3}a_{\rm M}]$. Panel (b) shows the maximum bulk current ${\cal J}$ associated with whirlpools and introduced in Eq.~(\ref{eq:novel_quantifier}) (in units of $\mu{\rm A}/{\rm nm}$). The direction of rotation of whirlpools is constant in $\psi \in [0,180\degree]$. By comparing panel (b) with panel (a), we conclude that ${\cal J}$ is a much better local proxy of topology compared to the average edge current ${\bar I}_{\|, {\rm edge}}$. Other parameters are set as following: $\theta =3.89\degree$, $w=-23.8~{\rm meV}$, and $W = 90\ \rm nm$.}
    \label{fig:edge_bulk}
\end{figure}

These facts are further investigated in Fig.~\ref{fig:edge_bulk}, where we plot two key quantities in the $(\psi,V)$ plane. In Fig.~\ref{fig:edge_bulk}(a) we plot the net current $\bar{I}_{\parallel, \rm edge}$ flowing at the bottom edge of the ribbon, which is defined by
\begin{equation}\label{eq:average_edge_current}
    \bar{I}_{\parallel, \rm edge} \equiv \frac{1}{a_{\rm M}} \int_0^{a_{\rm M}} dx \int_0^{\sqrt{3}a_{\rm M}} dy \ J_x(\bm{r})~.
\end{equation}
In Fig.~\ref{fig:edge_bulk}(b) we plot the quantity ${\cal J}$, which has been introduced in Eq.~(\ref{eq:novel_quantifier}).  
We have checked that ${\cal J}$ is numerically converged with respect to the transverse width $W$ of the ribbon. In both panels, the black dashed line corresponds to the transition between trivial $(\mathcal{C} = 0$ on the left) and topological $(\mathcal{C} = 1$ on the right) phases. 

Our first surprising finding is that the chirality of the ground-state edge currents, identified by the sign of $\bar{I}_{\parallel, \rm edge}$ in Fig.~\ref{fig:edge_bulk}(a), is {\it not} directly linked to the value of the Chern number.
In particular, in the trivial phase $\bar{I}_{\parallel, \rm edge}$ is approximately zero, as indicated by the white region. This is consistent with the current pattern shown in Fig.~\ref{fig:gs_currents}(a), corresponding to the black square in Fig.~\ref{fig:edge_bulk}(a).
In contrast, in the topological phase the edge current can be either right‑moving ($\bar{I}_{\parallel, \rm edge} > 0$, red region) or left‑moving ($\bar{I}_{\parallel, \rm edge} < 0$, blue region). The transition between these two regimes, where the edge current vanishes (white region), occurs entirely within the topological phase.
These opposite propagation directions can also be qualitatively inferred from Fig.~\ref{fig:gs_currents}(b) and (c), corresponding to the black triangle and black circle in Fig.~\ref{fig:edge_bulk}(a), respectively.

The second remarkable finding is that the presence of an intense whirlpool-like current flow in the bulk of the ribbon appears to be in a rather intimate relationship with the bulk-band topology, as seen in Fig.~\ref{fig:edge_bulk}(b). In this figure, the colored region, which denotes the region of parameter space where ${\cal J} \neq 0$, overlaps remarkably well with the topologically non-trivial region of the TPD. Moreover, the chirality of such whirlpools is constant across the entire $\mathcal{C} = 1$ phase and is reversed for $\psi > 180\degree$, where $\mathcal{C} = -1$. 

What is the reason behind the intimate link between $\bm J(\bm r)$ and bulk-band topology displayed in Fig.~\ref{fig:edge_bulk}(b)? As demonstrated in Sect.~\ref{sect:dispersive_geometric_contributions} of the SM~\cite{SM}, in a periodic system the ground-state current can be written as ${\bm J}({\bm r}) ={\bm J}_{\rm intra}({\bm r})+{\bm J}_{\rm inter}({\bm r})$ with
\begin{align}
    \bm J_{\rm intra}(\bm r) & =  \frac{e}{\hbar} \sum_{(n, {\bm k})\in {\rm occ.}} \rho_{nn; {\bm k}}(\bm r) \nabla_{\bm k} \varepsilon_n(\bm k) \ ,\label{eq:intraband_GS_current} \\
    \bm J_{\rm inter}(\bm r) & = -\frac{e}{\hbar} \sum_{(n, {\bm k}) \in {\rm occ.}} \sum_{m \neq n} (\varepsilon_{n}(\bm{k}) - \varepsilon_{m}(\bm{k}))\nonumber\\
    &\times {\rm Im} \Big [ \rho_{mn; {\bm k}}(\bm r)\bm{\mathcal{A}}_{nm}(\bm k)  \Big ]\ .\label{eq:interband_GS_current}
\end{align}
Here, $\varepsilon_n(\bm k)$ are Bloch-band energies and the quantities $\rho_{mn; {\bm k}}(\bm r)$ and $\bm{\mathcal{A}}_{nm}(\bm k)$, which depend on the periodic part $u_{{\bm k}, n}({\bm r})$ of the Bloch wave function, have been introduced in Sect.~\ref{sect:dispersive_geometric_contributions} of the SM~\cite{SM}. The quantity ${\bm J}_{\rm intra}(\bm r)$ is an intra-band contribution, which contains the semiclassical group velocity $\hbar^{-1}\nabla_{\bm{k}} \varepsilon_n(\bm{k})$. The inter-band contribution ${\bm J}_{\rm inter}(\bm r)$ entails quantum geometrical effects. Inter-band contributions are standard in Kubo formulas for linear-response functions~\cite{blount_1962, kruchkov_2023, verma_2025, Yu_2025}. (An example is reported below in Eq.~(\ref{eq:local_cond}) of the End Matter.) Moreover, they are known to generate ground-state current patterns following the spontaneous breaking of TRS occurring in the case of excitonic insulators~\cite{mazza_prbl_2023,mazza_arxiv_2026}.

Eqs.~(\ref{eq:intraband_GS_current})-(\ref{eq:interband_GS_current}) show that inter-band contributions do appear (and play a central role in systems with flat bands) also in a local equilibrium quantity. As the proof in Sect.~\ref{sect:dispersive_geometric_contributions} of the SM~\cite{SM} shows, it is precisely the fact that we focus on a ${\bm r}$-local quantity that unlocks the existence of the inter-band term (\ref{eq:interband_GS_current}). Integrating the ground-state current density ${\bm J}({\bm r})$ over space yields indeed zero because of the Bloch-Bohm theorem~\cite{bohm_1949}.

Because the bands of skyrmion Chern models~\cite{Wu2019,morales-duran_PRL_2024} are nearly flat, we find that the intra-band contribution (\ref{eq:intraband_GS_current}) is small and remains almost constant throughout the TPD. Nevertheless, it is useful to examine how it compares with the geometric inter-band contribution (\ref{eq:interband_GS_current}) in determining the total current. Referring to Fig.~\ref{fig:edge_bulk}(b), for $\psi = 170\degree$ (black circle) the intra-band term accounts for only $\approx 8\%$ of $\mathcal J$, while for $\psi = 107.7\degree$ (black triangle) it contributes approximately $\approx 5\%$. By contrast, for $\psi = 60\degree$, the local current is almost entirely of intra-band origin, with this contribution amounting to $\approx 90\%$ of the total.
Thus, within the topological phase, the ground-state current is dominated by the inter-band contribution, whereas in the topologically trivial phase the intra-band contribution prevails. This explains the accuracy of ${\cal J}$ as a proxy of topology, as noticed in Fig.~\ref{fig:edge_bulk}(b). In moir\'e materials, the flatness of the bands suppresses the trivial intra-band contribution to the local ground-state current, thereby allowing the geometric contribution to dominate.

In Sect.~\ref{sect:magnetic_field_whirlpools} of the SM~\cite{SM} we present numerical results concerning the magnetic field generated by the ground-state current pattern, which could be in principle measured by using NV-center magnetometry. From the Biot–Savart equations, we find field amplitudes in the range $\sim 1$–$10~\mu{\rm T}$ at nanometric distances from the ribbon. For probe distances $d \lesssim a_{\rm M}$, the magnetic field resolves the current structure inside the unit-cell, while for $d \gtrsim a_{\rm M}$, it retains only long-wavelength features, with a finite out-of-plane component in the bulk arising from large-scale current whirlpools.

Finally, in the End Matter we analyze the effect of the finite ribbon geometry on the Hall conductivity $\sigma_{xy}$. In a fully periodic 2D system, this ``global'' quantity is simply given by $\sigma^{({\rm bulk})}_{xy} = {\cal C} e^2/h$. Here, we investigate how this simple relation is approached as a function of the ribbon width $W$ and extract two ``critical exponents'' that govern its scaling behavior.

We hope that our results will stimulate further theoretical and experimental work on the spatial distribution of ground-state currents in moir\'e Chern insulators, possibly also in the fractional regime. In the future, it will also be interesting to study how these ground-state currents couple to (vacuum) electromagnetic fields~\cite{mazza_prbl_2023}.

{\it \color{blue} Acknowledgments.---}M.P. was supported by the European Union under grant agreement No. 101131579 - Exqiral. Views and opinions expressed are however those of the
author(s) only and do not necessarily reflect those of the European Union or the European Commission. Neither the European Union nor the granting authority can
be held responsible for them. F.T. acknowledges funding from MUR-PRIN 2022 - Grant No. 2022B9P8LN - (PE3)- Project NEThEQS ``Non-equilibrium coherent thermal effects in quantum systems” in PNRR Mission 4 - Component 2 - Investment 1.1 “Fondo per il Programma Nazionale di Ricerca e Progetti di Rilevante Interesse Nazionale (PRIN)'' funded by the European Union - Next Generation EU and the PNRR MUR project PE0000023-NQSTI, and from the Royal Society through the International Exchanges between the UK and Italy (Grant No. IEC R2 192166).

\clearpage
\section*{End Matter}
\subsection*{Finite size effects in the Hall conductivity}

The Hall conductivity $\sigma_{xy}$ is tied to the topological properties of a system's energy bands. It is indeed well known that in a periodic 2D system $\sigma_{xy}$ is quantized,
\begin{equation}\label{bulk_value}
\sigma^{({\rm bulk})}_{xy} = \mathcal{C}\frac{e^2}{h}~,
\end{equation}
$\mathcal{C}$ being the Chern number of the specified set of bands.  In a 2D ribbon of finite transverse width $W$, this simple result does not apply and finite-size effects are expected. We therefore study how the Hall conductivity of a twisted homobilayer TMD {\it ribbon} is affected by a finite $W$ in the topological phase and at the topological transition. In a ribbon, the Hall conductivity can be obtained via the following Kubo formula~\cite{GiulianiVignale}
\begin{eqnarray}\label{eq:local_cond} \nonumber
    &&\sigma_{xy}(\omega=0)= -\frac{i\hbar}{S}\sum_{m\neq n}\sum_{k_x}(f_{k_x,m}-f_{k_x,n}) \times \\
    &&\times \frac{\langle k_x, m \vert \ \hat{\jmath}_{x} \ \vert k_x,n \rangle\langle k_x, n \vert \ \hat{\jmath}_{y} \ \vert k_x,m \rangle}{(\varepsilon_{k_x,m}-\varepsilon_{k_x,n})(\varepsilon_{k_x, m}-\varepsilon_{k_x,n}+i0^+)}~,
\end{eqnarray}
where $S = a_{\rm M} W$ is the area of a rectangle, which, once periodically repeated along the $x$ direction, produces our ribbon geometry.
In Eq.~(\ref{eq:local_cond}), $\hat{\jmath}_{x}$ and $\hat{\jmath}_{y}$ are the $x$ and $y$ components of the current density operator $\hat{\bm{\jmath}}$, as obtained from the single-particle current density operator $\hat{\bm{\jmath}}(\bm r)$ introduced in the main text in Eq.~(\ref{jj}):
\begin{align}
    \hat{\bm{\jmath}} =\int_{S}  \hat{\bm{\jmath}}(\bm{r})\; d^2\bm r~.
\end{align}

\begin{figure}[h]
    \begin{overpic}[width=0.8\linewidth]{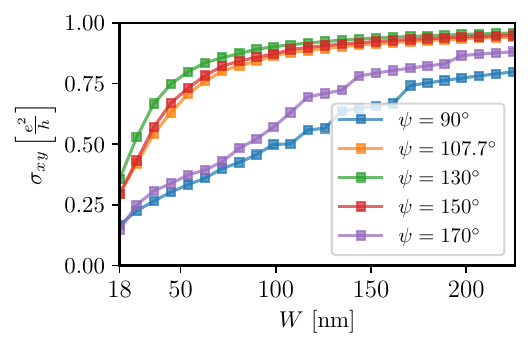}
        \put(2,60){(a)}
    \end{overpic}
    \begin{overpic}[width=0.85\linewidth]{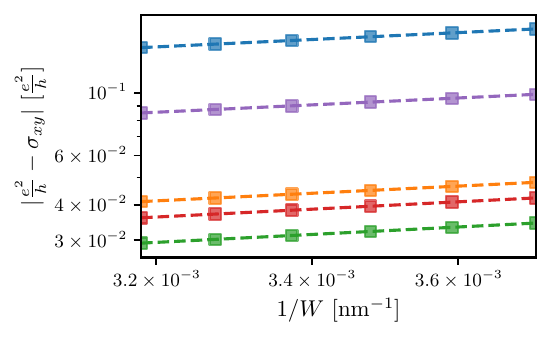}
        \put(6,60){(b)}
    \end{overpic}
    \caption{Dependence of the Hall conductivity $\sigma_{xy}$ of a twisted TMD ribbon on its width $W$, for various values of $\psi$. Data marked by orange squares refer to twisted MoTe$_2$. For all data displayed, the Chern number is ${\cal C}=+1$.
    Panel (a) shows that $\sigma_{xy}$ converges to ${e^2}/{h}$ for large widths. Panel (b) shows a log-log plot of the distance between $\sigma_{xy}$ and the bulk value ${e^2}/{h}$ as a function of $1/W$, for large values of $W$ (up to 35 rectangular unit cells). The dashed lines are power-law fit to the data. The color coding is the same as panel (a). All other parameters are fixed to $\theta = 3.89\degree$, $w=-23.8~{\rm meV}$, and $V=20.8~{\rm meV}$.}
    \label{fig:sigmaxy}
\end{figure}

In Fig.~\ref{fig:sigmaxy}(a), we show $\sigma_{xy}$ as a function of $W$ for several values of $\psi$, all in the ${\cal C} = 1$ phase. For $\psi$ values sufficiently far from the critical points $\psi_{\rm c}$ (approximately $85^\circ$ and $180^\circ$), the conductivity converges rapidly, reaching its asymptotic value already at widths of around 100~nm, as shown by the orange, green, and red curves. For $\psi = 90\degree$ and $\psi = 170\degree$, which are close to the phase boundaries, the convergence is much slower.

A more detailed analysis can be performed by examining the convergence of $\sigma_{xy}$ to $\sigma^{({\rm bulk})}_{xy}$ in the large-$W$ limit. In Fig.~\ref{fig:sigmaxy}(b), we present a log--log plot of the quantity $|\sigma_{xy} - \sigma^{({\rm bulk})}_{xy}|$ as a function of the inverse ribbon width $1/W$, focusing on sufficiently large values of $W$. Dashed lines represent fits of the numerical data obtained by using the following power-law decay:
\begin{equation}\label{eqn:scaling}
    \left|\sigma_{xy} - \sigma^{({\rm bulk})}_{xy}\right|\propto  W^{-\beta}~.
\end{equation}
Remarkably, we find that the fits are consistent with $\beta \approx 1$, independently of the value of $\psi$.
Another scaling law can be found. Indeed, by replacing the $x$-axis of Fig.~\ref{fig:sigmaxy}(b), originally given by $1/W$, with the scaling variable $(\psi - \psi_{\rm c})W^{1/\zeta}$, where $\psi_{\rm c}$ denotes the critical value of $\psi$ at the nearest phase boundary, all curves collapse onto a single universal curve for $\zeta \approx 1$.
This scaling behavior holds globally, i.e.~for all values of $\psi$ and $W$, and for both transition points $\psi_{\rm c}$, despite the fact that ${\cal C}$ changes from $0$ to $+1$ in one case and from $-1$ to $+1$ in the other.
Note that the two exponents, $\beta$ and $\zeta$, are independent. Additional results can be found in Sect.~\ref{sect:phase_transition_scaling_supp} of the SM~\cite{SM}.

A similar scaling analysis of topological properties was carried out in Ref.~\cite{caio_NATPHYS_2019} for the Haldane model. There, it was shown that the local Chern marker evaluated in the bulk can be used to characterize the critical behavior of topological phase transitions. Specifically, if $M$ denotes the control parameter and $M_{\rm c}$ its critical value, and $L$ is the linear system size (with open boundary conditions along both $x$ and $y$), the Chern marker was found to depend only on the scaling variable $(M - M_{\rm c})L^{1/\zeta}$, with $\zeta \approx 1$.
It is interesting to note that the same scaling law holds in our system for the Hall conductivity $\sigma_{xy}$ (a global measurable quantity compared to the local Chern marker) with $M$ replaced by $\psi$ and $L$ by $W$.

\clearpage
\onecolumngrid            
\setcounter{page}{1}

\setcounter{secnumdepth}{3} 
\setcounter{section}{0}
\setcounter{subsection}{0}
\setcounter{subsubsection}{0}
\setcounter{equation}{0}
\setcounter{figure}{0}
\setcounter{table}{0}

\makeatletter

\renewcommand{\thesection}{S\arabic{section}}
\renewcommand{\thesubsection}{S\arabic{section}.\arabic{subsection}}
\renewcommand{\thesubsubsection}{S\arabic{section}.\arabic{subsection}.\arabic{subsubsection}}

\renewcommand{\p@section}{}
\renewcommand{\p@subsection}{}
\renewcommand{\p@subsubsection}{}

\renewcommand{\theequation}{S\arabic{equation}}
\renewcommand{\thefigure}{S\arabic{figure}}
\renewcommand{\thetable}{S\arabic{table}}

\renewcommand{\p@equation}{}
\renewcommand{\p@figure}{}
\renewcommand{\p@table}{}

\@ifundefined{theHsection}{}{%
  \renewcommand{\theHsection}{S\arabic{section}}
  \renewcommand{\theHsubsection}{S\arabic{section}.\arabic{subsection}}
  \renewcommand{\theHsubsubsection}{S\arabic{section}.\arabic{subsection}.\arabic{subsubsection}}
  \renewcommand{\theHequation}{S\arabic{equation}}
  \renewcommand{\theHfigure}{S\arabic{figure}}
  \renewcommand{\theHtable}{S\arabic{table}}
}

\def\@seccntformat#1{\csname the#1\endcsname.\quad}
\makeatother

\renewcommand{\bibnumfmt}[1]{[S#1]}
\renewcommand{\citenumfont}[1]{S#1}

\begin{center}
\textbf{\Large Supplemental Material for:\\ ``Persistent currents, whirlpools, and local Chern markers in twisted TMD  Chern insulators''}

\bigskip

%
%
%
%
%
%
%

\bigskip

In this Supplemental Material we present details on the numerical method for finding the solution of the Schr\"{o}dinger's equation for a twisted TMD ribbon described by the Hamiltonian in Eq.~\eqref{H_up}, the definition of the current operator and how it can be divided into a dispersive and a geometric contribution, numerical results on the magnetic field generated by current whirlpools, as well as additional results concerning the scaling properties of $\sigma_{xy}$.

\end{center}

\onecolumngrid


\section{Solution of the Schrödinger equation for a  ribbon}
\label{eq:Schroed_ribbon_geometry}

We consider the continuum model Hamiltonian introduced in Ref.~\cite{Wu2019_supp}, acting in the layer pseudospin space:
\begin{equation}\label{eq:hamiltonian_supp}
\mathcal{H}_{\uparrow} = 
\begin{pmatrix}
-\frac{\hbar^2}{2m^*}(-i\bm{\nabla}-\bm{k}_{\rm b})^2 + \Delta_{\rm b}(\bm{r}) & \Delta_{\rm T}(\bm{r}) \\
\Delta_{\rm T}^*(\bm{r}) & -\frac{\hbar^2}{2m^*}(-i\bm{\nabla}-\bm{k}_{\rm t})^2 + \Delta_{\rm t}(\bm{r})\
\end{pmatrix}\ .
\end{equation}
For a ribbon geometry that is periodic along $x$ with period $a_{\rm M}$, we employ Bloch’s theorem and expand the wavefunction as
\begin{equation}
\Psi_{k_x,n}(\bm{r}) = e^{ik_x x} \sum_q e^{i \frac{2\pi}{a_{\rm M}} q x}
\begin{pmatrix}
u^{(\rm b)}_{k_x,n}(q,y) \\[0.4em]
u^{(\rm t)}_{k_x,n}(q,y)
\end{pmatrix} \ .
\end{equation}
The moiré potentials couple different Fourier components $q$, resulting in a set of coupled 1D equations along $y$.

In terms of exponential harmonics, the intralayer and interlayer terms take the form
\begin{align}
\Delta_{\rm b/t}(\bm{r}) &= 2V\Big[\cos(G_0 y \pm \psi) 
+ \cos\!\left(\tfrac{G_0}{2}y \mp \psi\right)(e^{i\frac{2\pi}{a_{\rm M}}x} + e^{-i\frac{2\pi}{a_{\rm M}}x})\Big] \ , \\
\Delta_{\rm T}(\bm{r}) &= w\Big[1 + 2\cos\!\left(\tfrac{G_0}{2}y\right)e^{-i\frac{2\pi}{a_{\rm M}}x}\Big] \ ,
\end{align}
where $G_0 = \frac{4\pi}{\sqrt{3}a_{\rm M}}$. These terms induce couplings between $q$ and $q\pm1$.

\begin{figure*}[h]\label{fig:t_bands}
    \centering
    \begin{overpic}
        [width = \textwidth]{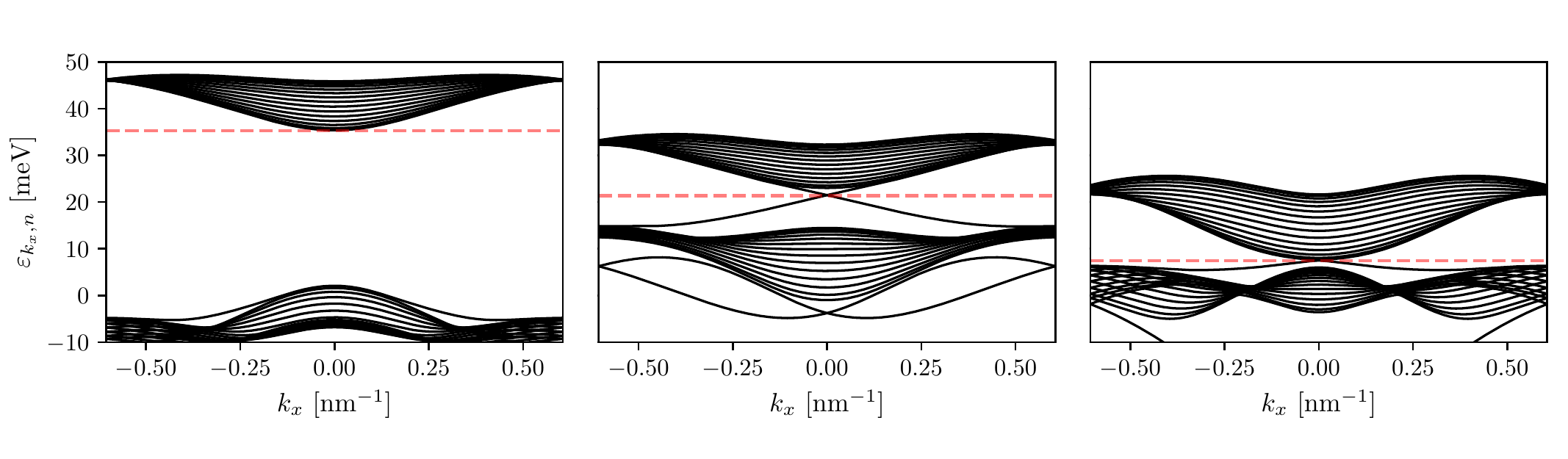}
        \put(5,27){(a)}
        \put(38,27){(b)}
        \put(69,27){(c)}
    \end{overpic}
    \caption{Transverse bands of the ribbon for the three cases considered in Fig. \ref{fig:gs_currents} of the main text. The red dashed line corresponds to the Fermi energy chosen in the three cases. (a) $\psi = 60\degree$, (b) $\psi = 107.7\degree$ and (c) $\psi = 170\degree$. Other parameters are set as follows: $\theta =3.89\degree$, $w=-23.8~{\rm meV}$, $V=20.8~{\rm meV}$, and $W = 90\ \rm nm$.}
\end{figure*}

After projecting onto a given Fourier component and performing the gauge transformation

\begin{equation}
u^{(\gamma)}_{k_x,n}(q,y) = e^{i\kappa_y y}\,\tilde{u}^{(\gamma)}_{k_x,n}(q,y) \ ,
\end{equation}
 with $\gamma = {\rm b}, {\rm t}$ and $\kappa_y = \frac{2\pi}{\sqrt{3}a_{\rm M}}$,  the problem reduces to coupled differential equations for $\tilde{u}^{(\gamma)}_{k_x,n}(q,y)$. Introducing the spinor
\begin{equation}
\tilde{\bm{u}}_{k_x,n}(q,y) =
\begin{pmatrix}
\tilde{u}^{(\rm b)}_{k_x,n}(q,y) \\
\tilde{u}^{(\rm t)}_{k_x,n}(q,y)
\end{pmatrix} \ ,
\end{equation}
the eigenvalue problem can be written compactly as
\begin{equation}
\hat{H} \tilde{\bm{u}}_{k_x,n}(q,y) = \varepsilon_{k_x,n} \tilde{\bm{u}}_{k_x,n}(q,y)\ ,
\end{equation}
with
\begin{equation}
\hat{H} = \hat{H}_{\rm kin} + \hat{H}_{\rm intra} + \hat{H}_{\rm inter} \ .
\end{equation}
The kinetic term reads
\begin{equation}
\hat{H}_{\rm kin} = -\frac{\hbar^2}{2m}
\left\{\left[k_x + \frac{2\pi}{a_{\rm M}}q + \kappa_x \sigma_z \right]^2 - \partial_y^2 \right\}\ ,
\end{equation}
where $\kappa_x = \frac{2\pi}{3a_{\rm M}}$. The intralayer and interlayer terms encode, respectively, the mixing of different momenta $q$ and layer hybridization,
\begin{align}
\hat{H}_{\rm intra} &= 2V\Big[\cos(G_0 y)\cos\psi\,\openone - \sin(G_0 y)\sin\psi\,\sigma_z \Big] \nonumber \\
&\quad + 2V\Big[\cos\!\left(\tfrac{G_0}{2}y\right)\cos\psi\,\openone + \sin\!\left(\tfrac{G_0}{2}y\right)\sin\psi\,\sigma_z \Big](\hat S_+ + \hat S_-) \ , \\
\hat{H}_{\rm inter} &= w\,\sigma_x + 2w \cos\!\left(\tfrac{G_0}{2}y\right)
\left(\sigma_+ \hat S_+ +  \sigma_- \hat S_- \right) \ ,
\end{align}
where $\hat S_\pm$ are matrices acting in momentum space which shift the index $q \to q \pm 1$, accounting for the mixing of different Fourier modes induced by the periodic potential. These are infinite matrices with elements $[ \hat S_\pm ]_{i,j} = \delta_{i \pm 1, j}$, acting as raising and lowering operators in Fourier space. In order to solve the eigenvalue problem these matrices must be truncated at a cutoff momentum $\vert q \vert \leq q_{\rm cutoff}$. The value of $q_{\rm cutoff}$ is then increased until convergence of the result is achieved.
Finally, we discretize real-space operators along the $y$-direction, by mapping $\partial_y^2$ onto the usual finite-difference operator:
\begin{equation}
\partial_y^2 \tilde{\bm{u}}(y_i) \simeq \frac{\tilde{\bm{u}}(y_{i+1}) - 2\tilde{\bm{u}}(y_i) + \tilde{\bm{u}}(y_{i-1})}{\Delta y^2}\ .
\end{equation}
where $\Delta y$ is the spacing of the numerical grid used for the discretization. In order for the solution to be accurate, we need $\Delta y \ll a_{\rm M}$, i.e. the discretization must be able to resolve the moiré potential. In this way, $\hat{H}$ becomes a sparse matrix in the $(\gamma, q,y)$ space, which can be diagonalized numerically. Results for the transverse bands are shown in Fig.~\ref{fig:t_bands}.

\section{Details on the current operator}
\label{sect:currop_SM}

The current operator is defined as $\hat{\bm{\jmath}}(\bm{r}) = -c\frac{\delta \hat{\mathcal{H}}_\uparrow(\bm{A}(\hat{\bm{r}}))}{\delta \bm{A}(\hat{\bm{r}})}$, where $\hat{\mathcal{H}}_\uparrow(\bm{A}(\hat{\bm{r}}))$ is the Hamiltonian in Eq.~\eqref{H_up}, minimally coupled to a vector potential $\bm{A}(\hat{\bm{r}})$:
\begin{equation}
    \hat{\mathcal{H}}_\uparrow(\bm{A}(\hat{\bm{r}}))= \begin{pmatrix} -\frac{\hbar^2}{2m^*}(\bm{k} - \bm{k}_{\rm b} + \frac{e}{c}\bm{A}(\hat{\bm{r}}))^2 + \Delta_{\rm b}(\bm{r}) & \Delta_{\rm T}(\bm{r}) \\ \Delta_{\rm T}^*(\bm{r}) & -\frac{\hbar^2}{2m^*}(\bm{k} - \bm{k}_{\rm t} + \frac{e}{c}\bm{A}(\hat{\bm{r}}))^2 + \Delta_{\rm t}(\bm{r}) \end{pmatrix} \ .
\end{equation}
By computing the functional derivative, we find:
\begin{equation}
    \hat{\bm{\jmath}}(\bm{r}) = \sum_i \hat{\bm{\jmath}}_i(\bm{r}) = \sum_i \begin{pmatrix} \hat{\bm{\jmath}}^{\rm (b)}_i(\bm{r}) & 0 \\ 0 & \hat{\bm{\jmath}}^{\rm (t)}_i(\bm{r}) \end{pmatrix} \ .
\end{equation}
Here, we introduced the single-electron current operators acting on the bottom and top layers $\hat{\bm{\jmath}}^{\rm (b/t)}_i(\bm{r})$:
\begin{equation}
\hat{\bm{\jmath}}^{\rm (b/t)}_i(\bm{r}) = \frac{e\hbar}{2m^*}\Big \{\hat{\bm{k}}_i - \bm{k}_{\rm b/t}, \delta(\bm{r}-\hat{\bm{r}}_i)\Big\} + \frac{e^2}{m^* c} \bm{A}(\hat{\bm{r}}_i)\delta(\bm{r}-\hat{\bm{r}}_i) \ ,
\end{equation}
where $\{\cdot,\cdot\}$ is the usual anti-commutator. Since we are interested in the ground-state current in the absence of external fields, we set $\bm{A}(\bm{r}) = \bm 0$ from now on.

The ground state of the system is a Slater determinant of the single-particle eigenstates in Eq.\eqref{eqn:H_tot}. The current operator $\hat{\bm{\jmath}}(\bm{r})$ is a sum of one-body operators acting on each electron. Since the Slater determinant is made up of orthonormal states, we simply need to evaluate the current carried by a single electron and then sum over the occupied states:
\begin{equation}\label{eq:current_supp}
    \bm{J}(\bm{r}) \equiv \langle {\rm GS} \vert \ \hat{\bm{\jmath}}(\bm{r}) \ \vert {\rm GS} \rangle = \sum_{i \in {\rm occ}} \langle \psi_i \vert \ \hat{\bm{\jmath}}_i(\bm{r}) \ \vert \psi_i \rangle \ ,
\end{equation}
where $\vert {\rm GS} \rangle = \prod_{i \in {\rm occ}} \hat{c}^\dagger_i \vert 0 \rangle$ and the sum runs over the occupied single-particle states $\vert \psi_i \rangle$. As discussed in the main text and in Sect.~\ref{eq:Schroed_ribbon_geometry} above, the eigenstates are labeled by two quantum numbers: the band index $n$ and the wave vector $k_x$, which resides in the 1D Brillouin zone $\left[-\frac{\pi}{a_{\rm M}}, \frac{\pi}{a_{\rm M}}\right]$. The band index $n$ includes the transverse modes due to the finiteness of the ribbon along $y$. At zero temperature, the occupied states are all those with energy below the Fermi energy. In Fig.~\ref{fig:t_bands} we show the transverse bands of our ribbon for $\psi = 60\degree, 107.7 \degree$ and $170\degree$, i.e. the three choices considered in Fig. (1) of the main text. The red dashed line corresponds to the Fermi energy chosen in the three cases. 

Notice that, since the continuum Hamiltonian in Eq.~\eqref{eqn:H_tot} contains infinitely many valence bands, the sum in Eq.~\eqref{eq:current_supp} formally runs over an infinite number of occupied states. To evaluate this sum, we decompose the current operator into its two single-spin/single-valley contributions, $\hat{\bm{\jmath}}_i(\bm r) = \hat{\bm{\jmath}}_{i, \uparrow}(\bm r) + \hat{\bm{\jmath}}_{i, \downarrow}(\bm r)$. Time-reversal symmetry implies that the local current contribution from an occupied spin-up state in valley $K$ is exactly canceled by that of its time-reversal partner, a spin-down state in valley $K^\prime$. Therefore, in the absence of either spontaneous or externally-induced time-reversal-symmetry breaking, the two contributions cancel identically. By contrast, if an exchange field (stemming from strong electron-electron interactions) spontaneously breaks time-reversal symmetry by rigidly shifting the spin-up bands relative to the spin-down bands, some states may be occupied while their time-reversal partners are empty. These unpaired occupied states then give rise to a finite local current. Consequently, only this finite subset of occupied states contributes to the ground-state expectation value of the total current operator.

\subsection{Dispersive and geometric contributions}
\label{sect:dispersive_geometric_contributions}

We now discuss how geometric and non-geometric terms contribute to the ground-state local current density $\bm J(\bm r)$ in a 2D crystal. Given the Bloch wavefunctions of the crystal $\vert \psi_{\bm k, n} \rangle$, with $n$ the band index and $\bm k$ the crystal momentum, the current is given by:
\begin{equation}
    \bm J(\bm r) = \frac{e}{2} \sum_{(n, {\bm k})\in {\rm occ.}} \langle \psi_{\bm k, n} \vert \{ \hat{\bm v} , \delta(\hat{\bm r} - \bm r) \}  \vert \psi_{\bm k, n} \rangle = e \sum_{(n, {\bm k})\in {\rm occ.}} \ {\rm Re} \Big [ \langle \psi_{\bm k, n} \vert \hat{\bm v} \delta(\hat{\bm r} - \bm r) \vert \psi_{\bm k, n} \rangle \Big ]\ ,
\end{equation}
where $\hat{\bm r}$ is the position operator and $\hat{\bm v}=\frac{i}{h}[\hat{H}, \hat{\bm r}]$ the velocity operator. For our purpose it is convenient to use the periodic Bloch eigenstates defined in the unit cell, i.e. $\vert u_{\bm k, n} \rangle = e^{- i \bm k \cdot \hat{\bm r}} \vert \psi_{\bm k, n} \rangle$.  Assuming for simplicity that only one band (the $n$-th one) is fully occupied and using the completeness relation $\sum_{m}\vert u_{\bm k, m} \rangle\langle u_{\bm k, m}\vert = \openone$, we find

\begin{eqnarray}\nonumber
    \bm J(\bm r) &=& e \sum_{\bm k}\ {\rm Re} \Big [ \langle u_{\bm k, n} \vert \hat{\bm v}(\bm k ) \delta(\hat{\bm r} - \bm r) \vert u_{\bm k, n} \rangle \Big ] = \\ \nonumber
    &=& e \sum_m \sum_{\bm k}\ {\rm Re} \Big [ \langle u_{\bm k, n} \vert \hat{\bm v}(\bm k ) \vert u_{\bm k, m} \rangle \langle u_{\bm k, m} \vert \delta(\hat{\bm r} - \bm r) \vert u_{\bm k, n} \rangle \Big ] = \\
    &=&  e \sum_m \sum_{\bm k}\ {\rm Re} \Big [ \langle u_{\bm k, n} \vert \hat{\bm v}(\bm k ) \vert u_{\bm k, m} \rangle \langle u_{\bm k, m} \vert \bm r \rangle \langle \bm r \vert u_{\bm k, n} \rangle \Big ]\ , \label{eq:J_r}
\end{eqnarray}
where we have introduced the $\bm k$-dependent velocity operator:
\begin{equation}
    \hat{\bm v}(\bm k ) \equiv e^{i \bm k \cdot \hat{\bm r}} \hat{\bm v} e^{-i \bm k \cdot \hat{\bm r}} = \frac{1}{\hbar} \nabla_{\bm k} \hat{H}(\bm k)\ .
\end{equation}
From the last expression we deduce the following form for the matrix elements of the velocity operator:
\begin{equation}
    \langle u_{\bm k, n} \vert \hat{\bm v}(\bm k ) \vert u_{\bm k, m} \rangle =  \delta_{nm} \frac{1}{\hbar}\nabla_{\bm{k}} \varepsilon_n(\bm{k}) - \frac{1}{\hbar}\varepsilon_{nm}(\bm{k}) \langle u_{\bm{k}, n} \vert \nabla_{\bm k} u_{\bm{k},m} \rangle\ ,
\end{equation}
where $\varepsilon_{nm}(\bm{k}) = \varepsilon_{n}(\bm{k}) - \varepsilon_{m}(\bm{k})$ is the excitation energy between two bands evaluated at a given $\bm k$. The first term is an intra-band contribution, which contains the semiclassical group velocity $\frac{1}{\hbar}\nabla_{\bm{k}} \varepsilon_n(\bm{k})$. The second term, which is purely off-diagonal in the band index, contains the effects of wavefunction geometry. We thus identify two contributions to the current in Eq.~\eqref{eq:J_r}: a dispersive (intra-band) one, $\bm J_{\rm intra}(\bm r)$, and a geometric (inter-band) one, $\bm J_{\rm inter}(\bm r)$:
\begin{align}
    & \bm J_{\rm intra}(\bm r) =  \frac{e}{\hbar} \sum_{\bm k} \rho_{nn; {\bm k}}(\bm r) \nabla_{\bm k} \varepsilon_n(\bm k) \ , \\
    & \bm J_{\rm inter}(\bm r) = -\frac{e}{\hbar} \sum_{m \neq n} \sum_{\bm k} \varepsilon_{nm}(\bm{k})\ {\rm Im} \Big [ \rho_{mn; {\bm k}}(\bm r)\bm{\mathcal{A}}_{nm}(\bm k)  \Big ] \ ,
\end{align}
where $\rho_{mn; {\bm k}}(\bm r) \equiv \langle u_{\bm k, m} \vert \bm r \rangle \langle \bm r \vert u_{\bm k, n} \rangle$ and the multi-band Berry connection $\bm{\mathcal{A}}_{nm}(\bm k) \equiv \langle u_{\bm{k}, n} \vert i\nabla_{\bm k} u_{\bm{k}, m} \rangle$ \cite{vanderbilt_supp}. The total current in Eq.~\eqref{eq:J_r} is therefore written as $\bm J(\bm r) = {\bm J}_{\rm intra}(\bm r) + {\bm J}_{\rm inter}(\bm r)$.

\section{Magnetic field associated with whirlpools}
\label{sect:magnetic_field_whirlpools}

Magnetometry based on nitrogen-vacancy (NV) centers~\cite{casola_2018_supp} probes the local current density by measuring the magnetic field above a device and inverting the Biot–Savart relation, enabling vector-resolved, noninvasive imaging of current patterns. In electronic conductors, for example, wide-field and scanning NV modalities directly visualized viscous (Poiseuille) electron flow in graphene at room temperature, validating the existence of hydrodynamic transport~\cite{viscous_flow_review_supp,pellegrino_whirlpools_PRB_supp} via a quantitative current pattern reconstruction~\cite{ku_2020_supp}. 
At low temperatures, scanning NV probes have mapped supercurrent distributions and Josephson vortices in Josephson junctions, revealing bias-tunable ``hidden'' configurations that are inaccessible to transport alone~\cite{chen_2024_supp}. In this Section, we quantify the magnetic field $\bm B(\bm r,z)$ produced by ground-state currents in twisted MoTe$_2$.

A 2D current density $\bm J(\bm r,z=0)$, confined at $z = 0$, generates a magnetic field at $|z| > 0$ and $\bm B(\bm r,z) = \left(B_x(\bm r,z),B_y(\bm r,z),B_z(\bm r,z)\right)$, with
components given by the Biot-Savart law:
\begin{equation}\label{eqn:Bx}
B_x(\bm r, z)
= \frac{\mu_0}{2}\int d^{2}\bm r^\prime\,
{\cal K}_{xy} \left(\bm r- \bm r^\prime, z\right)
J_y(\bm r^\prime)~,
\end{equation}

\begin{equation}\label{eqn:By}
B_y(\bm r, z)
= -\frac{\mu_0}{2}\int d^{2}\bm r^\prime\,
{\cal K}_{xy} \left(\bm r- \bm r^\prime, z\right)
J_x(\bm r^\prime)~,
\end{equation}

\begin{equation}\label{eqn:Bz}
B_z(\bm r, z)
= \frac{\mu_0}{2}\int d^{2}\bm r^\prime\, \Big [
{\cal K}_{zx} \left(\bm r- \bm r^\prime, z\right)
J_x(\bm r^\prime) - {\cal K}_{zy} \left(\bm r- \bm r^\prime, z\right)
J_y(\bm r^\prime) \Big ]~.
\end{equation}

The kernels ${\cal K}_{xy}, {\cal K}_{zx}$ and ${\cal K}_{zy}$ appearing in Eqs.~\eqref{eqn:Bx}-\eqref{eqn:Bz} are given by:
\begin{eqnarray}
    {\cal K}_{xy}\left(\bm r- \bm r^\prime, z\right) &=&\frac{z}{2\pi\left[(\bm r - \bm r^\prime)^2 + z^2\right]^{3/2}}~, \\
    {\cal K}_{zx}\left(\bm r- \bm r^\prime, z\right) &=&\frac{y-y^\prime}{2\pi\left[(\bm r - \bm r^\prime)^2 + z^2\right]^{3/2}}~,\\
    {\cal K}_{zy}\left(\bm r- \bm r^\prime, z\right) &=&\frac{x-x^\prime}{2\pi\left[(\bm r - \bm r^\prime)^2 + z^2\right]^{3/2}}~.
\end{eqnarray}

The above integration extends over all space. In order to evaluate the magnetic field, it is useful to exploit the periodicity of the current density along the $x$-direction. We can rewrite the above integrals as:

\begin{eqnarray}
    B_x(\bm r, z)
&=& \frac{\mu_0}{2} \int_0^{a_{\rm M}} dx^\prime \int_0^{W} dy^\prime\,
{\cal Q}_{xy} \left(\bm r- \bm r^\prime, z\right)
J_y(\bm r^\prime)~, \\
B_x(\bm r, z)
&=& -\frac{\mu_0}{2} \int_0^{a_{\rm M}} dx^\prime \int_0^{W} dy^\prime\,
{\cal Q}_{xy} \left(\bm r- \bm r^\prime, z\right)
J_x(\bm r^\prime)~, \\
B_z(\bm r, z)
&=& \frac{\mu_0}{2}\int_0^{a_{\rm M}} dx^\prime \int_0^{W} dy^\prime\, \Big [
{\cal Q}_{zx} \left(\bm r- \bm r^\prime, z\right)
J_x(\bm r^\prime) - {\cal Q}_{zy} \left(\bm r- \bm r^\prime, z\right)
J_y(\bm r^\prime) \Big ]~,
\end{eqnarray}
where the integral is now over a single horizontal period, and we have introduced the modified kernels:
\begin{eqnarray}
    {\cal Q}_{ij} \left(\bm r- \bm r^\prime, z\right) = \sum_{n = -\infty}^{+\infty} {\cal K}_{ij} \left(x- x^\prime - na_{\rm M}, y - y^\prime, z\right)~.
\end{eqnarray}
The convergence of these integrals was checked numerically.

\begin{figure}
    \centering
    \begin{overpic}[width=0.49\linewidth]{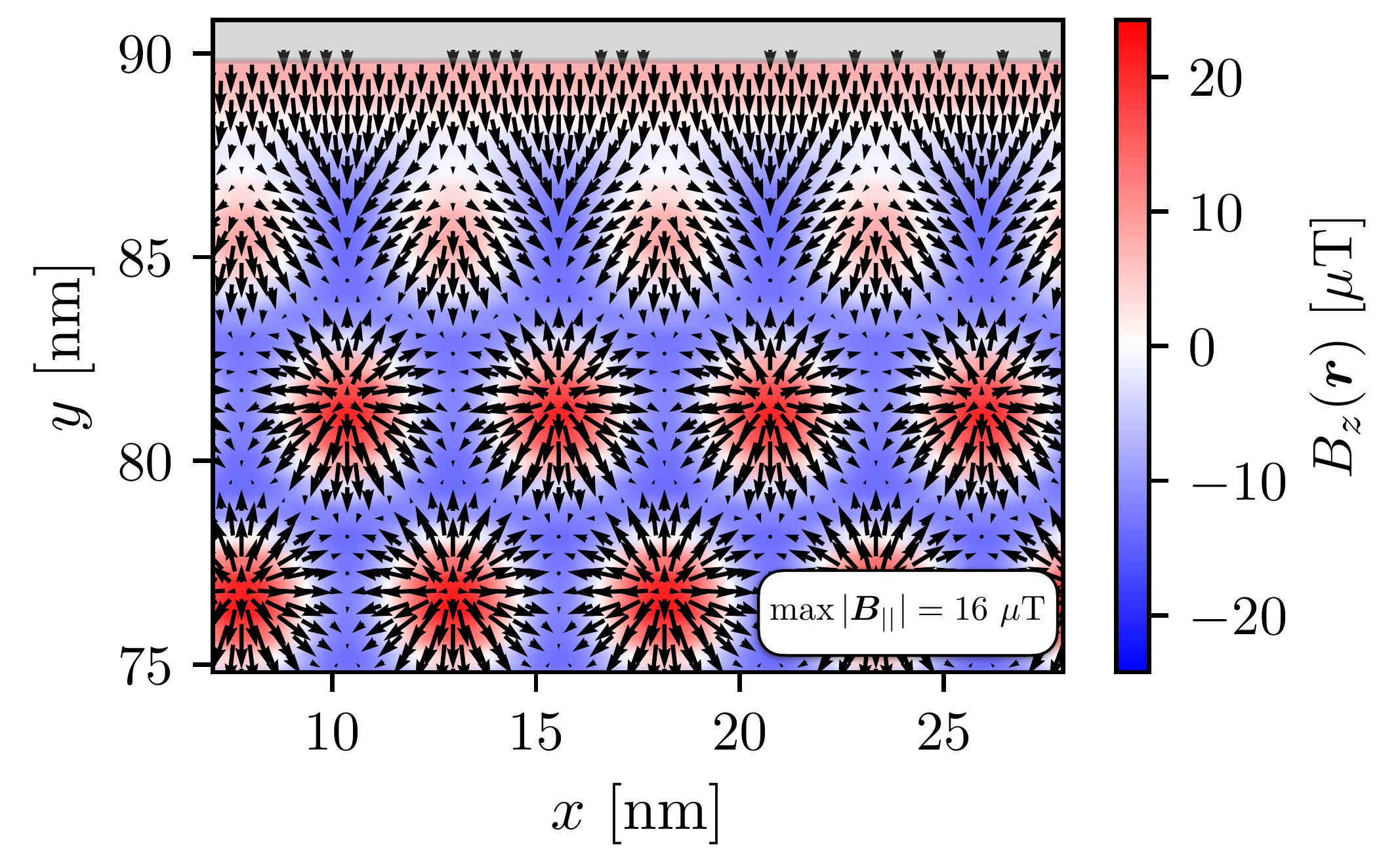}
        \put(10,65){(a)}
    \end{overpic}%
    \begin{overpic}[width=0.49\linewidth]{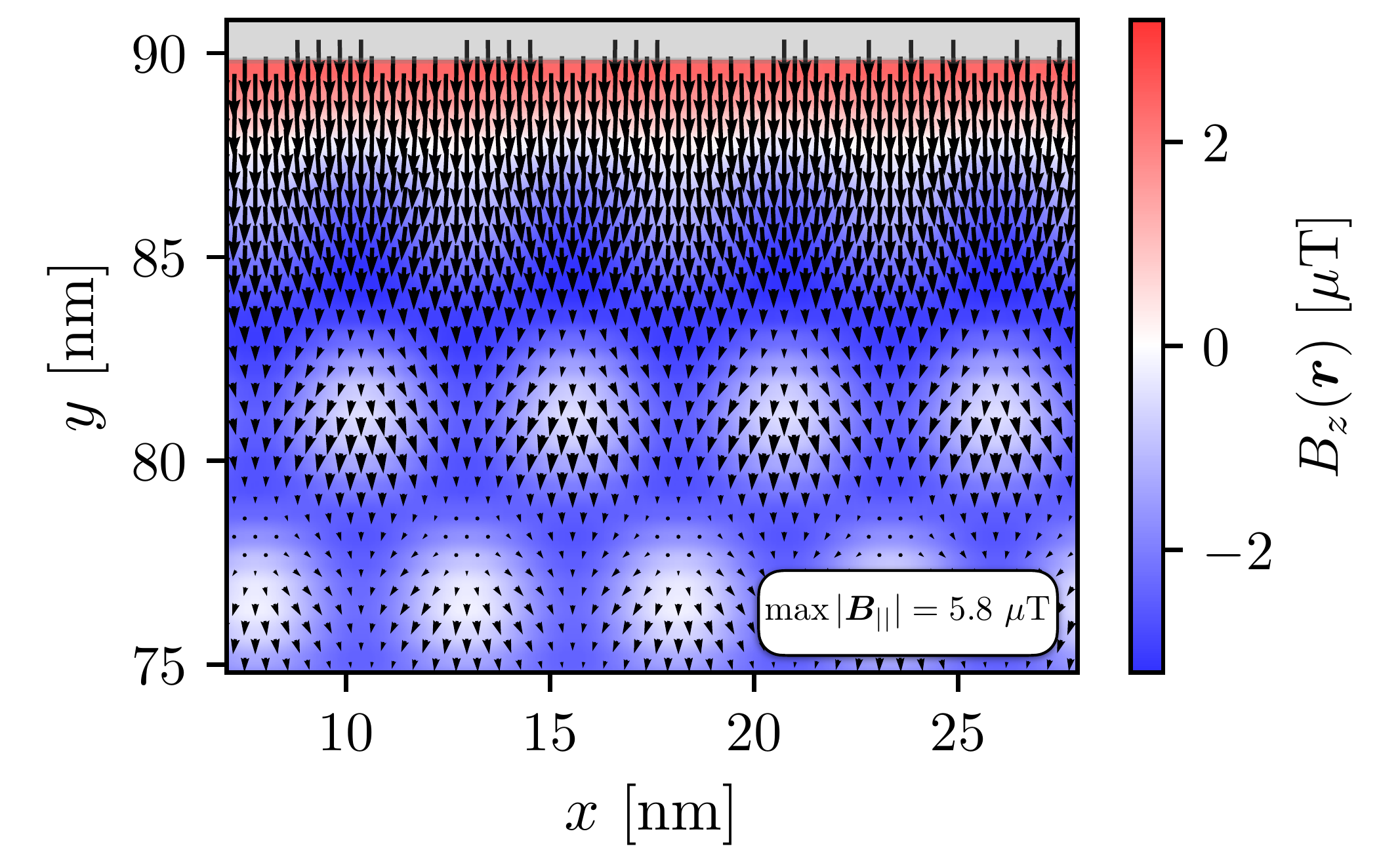}
        \put(10,65){(b)}
    \end{overpic}
    \begin{overpic}[width=0.49\linewidth]{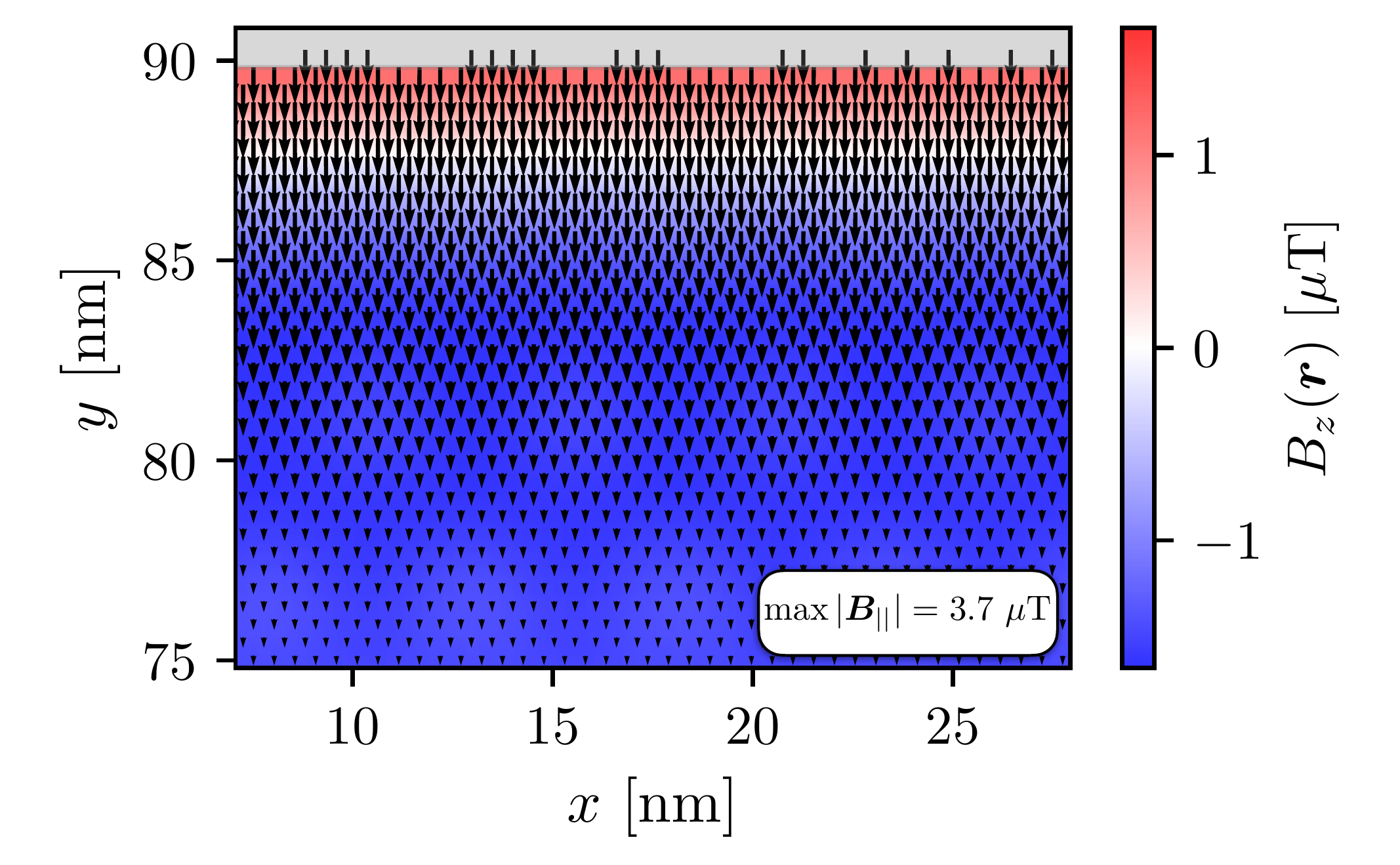}
        \put(10,65){(c)}
    \end{overpic}%
    \begin{overpic}[width=0.49\linewidth]{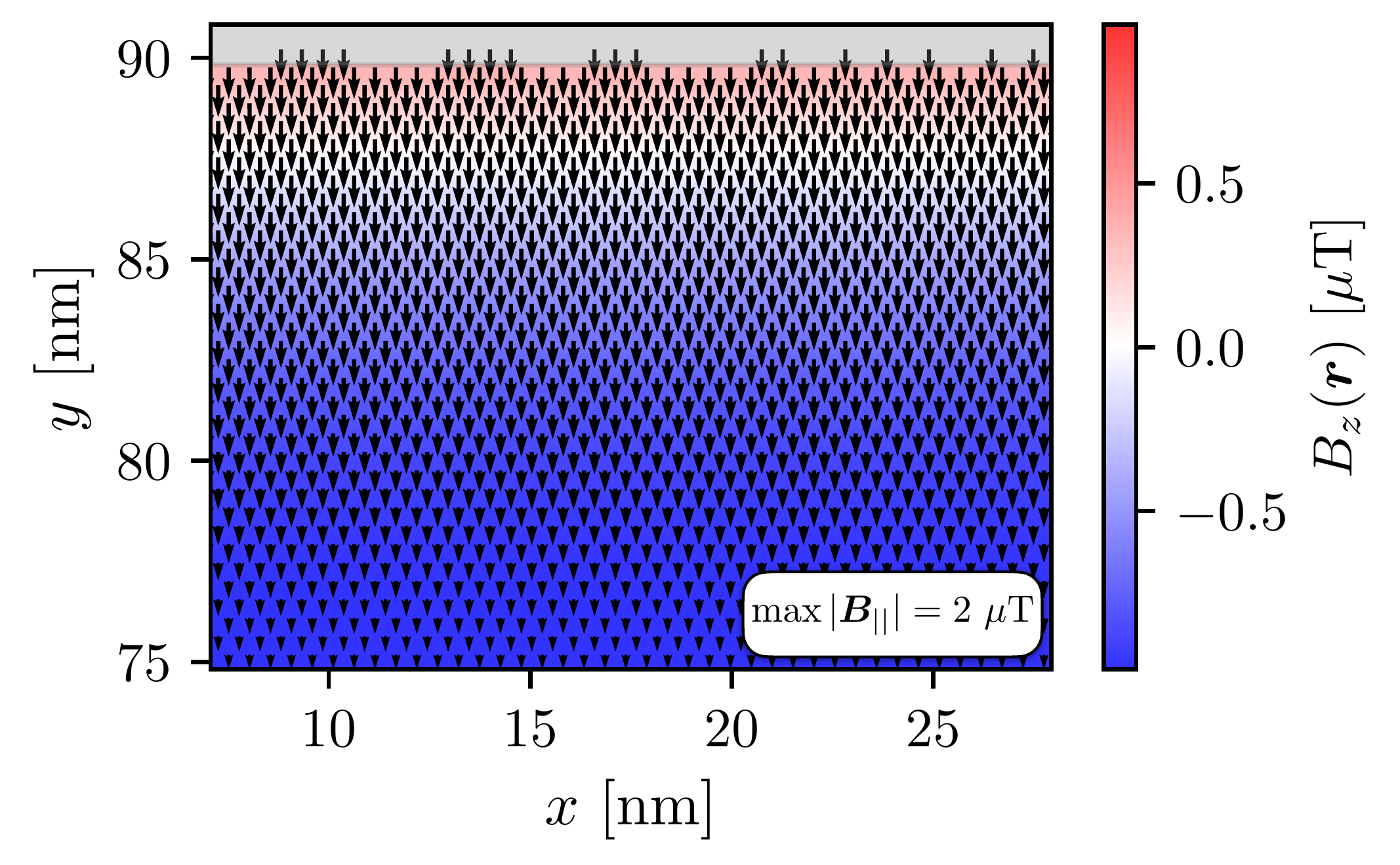}
        \put(10,65){(d)}
    \end{overpic}
    \caption{Magnetic field components $B_x(z), B_y(z)$ (vector field on the plane) and $B_z(z)$ (colormap), produced by the ground state currents of MoTe$_2$. For clarity, the maximum intensity of the parallel field is reported in the inset. Different panels refer to different values of $z$. Panel (a) $z = 1~{\rm nm}$; panel (b) $z = 3~{\rm nm}$; panel (c) $z = 5~{\rm nm}$; panel (d) $z = 10~{\rm nm}$. The phase $\psi$ is fixed to $107.7\degree$. All the other parameters are fixed as in Fig.~\ref{fig:t_bands}}
    \label{fig:B_vs_y}
\end{figure}

In Fig.~\ref{fig:B_vs_y}, we show the magnetic field components $B_x(z), B_y(z)$ as a vector field on the plane, and the component $B_z(z)$ as a color plot. This field is associated with the current patterns obtained in the near-critical phase (Fig. (1)-c in the main text) and is calculated for different values of the vertical distance $z = d$ from the ribbon plane, as obtained from Eqs.~\eqref{eqn:Bx}-\eqref{eqn:Bz}. The distance is set as $d = 1\ \rm nm$ [panel (a)], $d = 3\ \rm nm$ [panel (b)], $d = 5\ \rm nm$ [panel (c)] and $d = 10\ \rm nm$ [panel (d)]. For $d\lesssim a_{\rm M} \approx 5\ \rm nm$, all three components show significant variations on the scale of the unit cell, while for distances $d\gtrsim a_{\rm m}$, the magnetic field loses information about the current pattern within the unit cell but shows peaks in correspondence with the edge states. Most importantly, the $z$-component of the magnetic field has a finite value above the bulk of the system, even for $d\gtrsim a_{\rm m}$, due to the presence of large scale vortices.

\section{Analysis of the phase transition for varying $V$ at fixed $\psi$}
\label{sect:phase_transition_scaling_supp}

\begin{figure*}[h]
    \centering
    \begin{overpic}[width=0.5\linewidth]{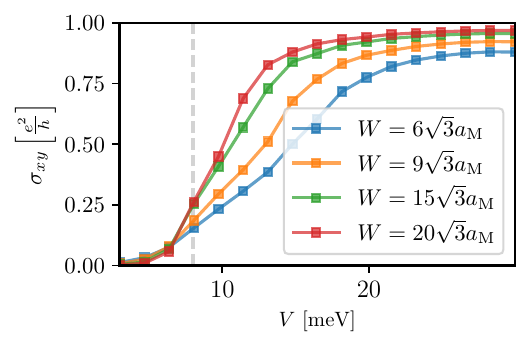}
        \put(0,60){(a)}
    \end{overpic}%
    \begin{overpic}[width=0.5\linewidth]{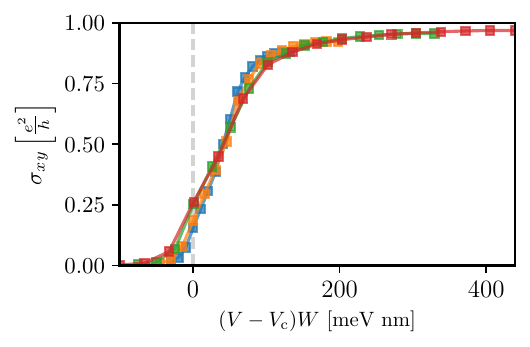}
        \put(0,60){(b)}
    \end{overpic}\\
    \begin{overpic}[width=0.5\linewidth]{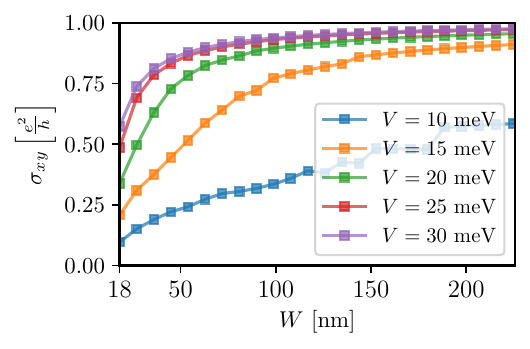}
        \put(0,60){(c)}
    \end{overpic}%
    \begin{overpic}[width=0.48\linewidth]{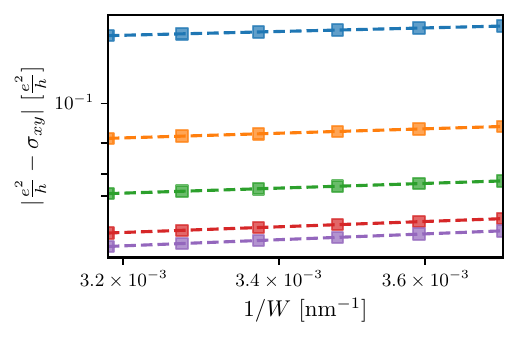}
        \put(10,60){(d)}
    \end{overpic}

    \caption{Hall conductivity $\sigma_{xy}$ of a ribbon of twisted TMD. Panel (a) shows $\sigma_{xy}$  as a function of the potential $V$ for a few values of the ribbon width $W$. Panel (b) is the same as panel (a) following a rescaling of the $x$-axis. When plotted as a function of $(V - V_{\rm c})W$, all the curves collapse on top of each other. In panel (c) the Hall conductivity $\sigma_{xy}$ is shown as a function of the ribbon width $W$ for few values of $V$. Panel (d) shows the distance of $\sigma_{xy}$ to the bulk value as a function of $1/W$, for large values of $W$ (i.e. 30-35 rectangular unit cells). The lines are power-law fit to the data and the results are shown in a log-log plot. The color coding is the same as panel (c). For each value displayed, the Chern number of the first valence band is ${\cal C}=1$. In all panels, the other parameters are fixed to $\theta = 3.89\degree$, $w=-23.8~{\rm meV}$, and $\psi = 140\degree$.}
    \label{fig:sigmaxy_supp}
\end{figure*}

In this Section we report a similar analysis of the Hall conductivity as done in the End Matter. In this case, we study the topological phase transition at fixed $\psi = 140\degree$, for several values of $V$. The results are shown in Fig. \ref{fig:sigmaxy_supp}, and are similar to the results of the analysis done at varying $\psi$. Most importantly, the same scaling laws hold: the transverse conductivity $\sigma_{xy}$ is found to be a function of $(V-V_{\rm c}) W^{1/\zeta}$ with $\zeta \approx 1$ and for large values of $W$ the conductivity approaches the bulk value with a power law dependence on the width, i.e. $\left|\sigma_{xy} - {\cal C}\frac{e^2}{h}\right|=\alpha  W^{-\beta}$ with $\beta \approx 1$.

\begin{thebibliography}{99}
\bibitem{foot_magnetization}
We stress that $\bm{J}(\bm{r})$ is a perfectly defined, unambiguous, local observable. This is not the case for the local orbital magnetization, which has physical meaning only as a macroscopically averaged quantity. See discussions in Refs.~\cite{Bianco_PRL_2013} and \cite{Hirst_1997}.
\bibitem{footnote_2}
Note that $\psi = 180\degree$ defines a high-symmetry line in the TPD, along which the local current density ${\bm J}({\bm r})$ is identically zero $\forall~V$.
\end{thebibliography}
\end{document}